\documentclass[sigconf]{acmart}

\usepackage{booktabs} 
\usepackage{ccicons}  
\usepackage{todonotes}
\usepackage{subfig}
\graphicspath{{figures/}}
\usepackage[english]{babel}
\usepackage[autostyle=true,english=american]{csquotes}
\usepackage{gensymb}
\usepackage{balance}
\usepackage[most]{tcolorbox}

\usepackage{acronym}
\newacro{AI}[AI]{Artificial Intelligence}
\newacro{UI}[UI]{user interface}
\newacro{GUI}[GUI]{graphical user interface}
\newacro{TLX}[TLX]{NASA-Task Load Index}
\newacro{RTLX}[Raw-TLX]{NASA Raw-Task Load Index}
\newacro{ER}[ER]{error rate}
\newacro{TCT}[TCT]{task completion time}
\newacro{HCI}[HCI]{Human-Computer Interaction}
\newacro{UX}[UX]{user experience}
\newacro{HFE}[HFE]{Human Factors and Ergonomics}
\newacro{cuDNN}[cuDNN]{CUDA Deep Neural Network library}
\newacro{RMSE}[RMSE]{root mean squared error}
\newacro{HMD}[HMD]{Head-Mounted Display}
\newacro{RF}[RF]{Random Forest}
\newacro{GP}[GP]{Gaussian process, long-plural = Gaussian processes}
\newacro{KNN}[\textit{k}NN]{\textit{k}-nearest neighbor}
\newacro{NN}[NN]{Neural Network}
\newacro{DNN}[DNN]{ Deep Neural Network}
\newacro{CNN}[CNN]{Convolutional Neural Network}
\newacro{FCL}[FCL]{fully connected layer}
\newacro{BoD}[BoD]{Back-of-Device}
\newacro{FOV}[FoV]{field of view}
\newacro{RW}[RW]{Real World}
\newacro{IFRC}[IFRC]{index finger ray cast}
\newacro{FRC}[FRC]{forearm ray cast}
\newacro{EFRC}[EFRC]{eye-finger ray cast}
\newacro{HRC}[HRC]{Human-Robot Collaboration}
\newacro{HRI}[HRI]{Human-Robot Interaction}
\newacro{6DOF}[6DOF]{six-degree-of-freedom}
\newacro{LOOCV}[LOOCV]{leave-one-out cross-validation}
\newacro{CV}[CV]{cross-validation}
\newacro{RM}[RM]{repeated measure}
\newacro{ANOVA}[ANOVA]{analysis of variance}
\newacro{RMANOVA}[RM-ANOVA]{repeated measures analysis of variance}
\newacro{AGATe}[AGATe]{AGreement Analysis Toolkit}
\newacro{GHoST}[GHoST]{Gesture Heatmap Toolkit Gesture Heatmaps Toolkit}
\newacro{GREAT}[GREAT]{Gesture Relative Accuracy Toolkit}
\newacro{GRT}[GRT]{Gesture Recognition Toolkit}
\newacro{DTW}[DTW]{Dynamic Time Warping}
\newacro{LHRD}[LHRD]{large high resolution display}
\newacro{GEQ}[GEQ]{Game Experience Questionnaire}
\newacro{SPGQ}[SPGQ]{Social Presence Gaming Questionnaire}
\newacro{JND}[JND]{just-noticeable difference}
\newacro{SUS}[SUS]{system usability scale}
\newacro{CSCW}[CSCW]{computer-supported cooperative work}
\newacro{CAD}[CAD]{computer-aided design}
\newacro{MR}[MR]{Mixed Reality}
\newacro{CVE}[CVE]{Collaborative Virtual Environment}
\newacro{AR}[AR]{Augmented Reality}
\newacro{AV}[AV]{Augmented Virtuality}
\newacro{VR}[VR]{Virtual Reality}
\newacro{PRISMA}[PRISMA]{Preferred Reporting Items for Systematic Reviews}
\newacro{PRISMA-Scope}[PRISMA-ScR]{Meta-Analyses Extension for Scoping Reviews}
\newacro{TF-IDF}[TF-IDF]{Term Frequency-Inverse Document Frequency}
\newacro{TF}[TF]{Term Frequency}
\newacro{AVs}[AVs]{Automated Vehicles}
\newacro{eHMIs}[eHMIs]{external Human-machine interfaces}
\newacro{SAR}[SAR]{Spatial Augmented Reality}
\newacro{IFR}[IFR]{International Federation of Robotics}
\newacro{ADL}[ADL]{Activity of Daily Living} \newacroplural{ADL}[ADLs]{Activities of Daily Living}
\newacro{LED}[LED]{Light-Emitting Diode}
\newacro{DoF}[DoF]{Degree-of-Freedom} \newacroplural{DoF}[DoFs]{Degrees-of-Freedom}
\newacro{HHC}[HHC]{Human-Human Collaboration}
\newacro{IDF}[IDF]{Inverse Document Frequency}
\newacro{DnD}[D\&D]{Design and Development}
\newacro{XR}[XR]{Extended Reality}
\newacro{BCI}[BCI]{Brain-computer Interfaces}
\newacro{ADMC}[ADMC]{Adaptive DoF Mapping Control}
\newacro{IMU}[IMU]{Inertial Measurement Unit}
\newacro{QUEAD}[QUEAD]{Questionnaire for the Evaluation of Physical Assistive Devices}
\newacro{ROS}[ROS]{Robot Operating System}
\newacro{LAN}[LAN]{Local Area Network}

\AtBeginDocument{%
  \providecommand\BibTeX{{%
    \normalfont B\kern-0.5em{\scshape i\kern-0.25em b}\kern-0.8em\TeX}}}

\copyrightyear{2024} 
\acmYear{2024} 
\setcopyright{rightsretained} 
\acmConference[HRI '24 Companion]{Companion of the 2024 ACM/IEEE International Conference on Human-Robot Interaction}{March 11--14, 2024}{Boulder, CO, USA}
\acmBooktitle{Companion of the 2024 ACM/IEEE International Conference on Human-Robot Interaction (HRI '24 Companion), March 11--14, 2024, Boulder, CO, USA}
\acmDOI{10.1145/3610978.3640626}
\acmISBN{979-8-4007-0323-2/24/03}

\begin{document}

\title{Exploring of Discrete and Continuous Input Control for AI-enhanced Assistive Robotic Arms}

\author{Max Pascher}
\orcid{0000-0002-6847-0696}
\email{max.pascher@udo.edu}
\affiliation{
    \institution{TU Dortmund University}
    \city{Dortmund}
    \country{Germany}
}
\affiliation{
    \institution{University of Duisburg-Essen}
    \city{Essen}
    \country{Germany}
}

\author{Kevin Zinta}
\orcid{0009-0005-1005-1885}
\email{kevin.zinta@studmail.w-hs.de}
\affiliation{
    \institution{Westphalian University of Applied Sciences}
    \city{Gelsenkirchen}
    \country{Germany}
}



\author{Jens Gerken}
\orcid{0000-0002-0634-3931}
\email{jens.gerken@udo.edu}
\affiliation{
    \institution{TU Dortmund University}
    \city{Dortmund}
    \country{Germany}
}

\renewcommand{\shortauthors}{Max Pascher, Kevin Zinta, and Jens Gerken}

\begin{abstract}

Robotic arms, integral in domestic care for individuals with motor impairments, enable them to perform \acp{ADL} independently, reducing dependence on human caregivers. These collaborative robots require users to manage multiple \acp{DoF} for tasks like grasping and manipulating objects. Conventional input devices, typically limited to two \acp{DoF}, necessitate frequent and complex mode switches to control individual \acp{DoF}. Modern adaptive controls with feed-forward multi-modal feedback reduce the overall task completion time, number of mode switches, and cognitive load. Despite the variety of input devices available, their effectiveness in adaptive settings with assistive robotics has yet to be thoroughly assessed. This study explores three different input devices by integrating them into an established \acs{XR} framework for assistive robotics, evaluating them and providing empirical insights through a preliminary study for future developments.


\end{abstract}




\maketitle

\section{Introduction}
The progress in the development of (semi-)autonomous technologies compelled their incorporation into numerous sectors, reshaping how we live and work. This integration includes scenarios of close collaboration with robotic devices, ranging from industrial assembly lines~\cite{Braganca.2019} to personal mobility aids~\cite{Fattal.2019}. Among these collaborative technologies, assistive robotic arms emerge as a particularly valuable and versatile subset, finding applications across various domains (e.g.,~\cite{Pulikottil.2021,Beaudoin.2018}).


Assistive robotic arms can enhance the independence of individuals with restricted mobility~\cite{Kyrarini.2021survey,Pascher.2021recommendations}. These technologies -- particularly when integrated with \ac{AI} -- empower individuals to perform \acfp{ADL}, which often entail tasks like gripping and manipulating objects in their surroundings, without reliance on human assistance~\cite{Petrich.2022ADL}. However, current \ac{HRI} research underscores a notable challenge faced by developers: optimizing the autonomy level of assistive robots~\cite{Lebrasseur.2019}. Striking a balance is crucial, as purely autonomous systems may diminish user interaction and trust, while manual controls could prove impractical for users with specific impairments~\cite{Pollak.2020,Kim.2012,zlotowski2017can}. 
Shared control -- combining manual input with algorithmic assistance -- emerges as such a balanced approach and a promising research direction.  


In this work, we explore three different input devices for controlling an assistive robotic arms in shared control applications: 
\begin{itemize}
    \item \emph{Joy-Con}: A motion controller with continuous data input, suited for one-handed operation. 
    \item \emph{Head}: User control input by head-based movements, using continuous data. 
    \item \emph{Button}: A set of assistive buttons to control the robot in an accessible manner with discrete input data.
\end{itemize}





\section{Related Work}

Standard control devices with a high \acf{DoF}, like gaming joysticks and keyboards, often pose challenges for users with severe motor impairments. Addressing these issues requires alternative solutions, such as specialized training or different interfaces~\cite{Peralta.2007,Taheri2021Design}.
An approach proposed by \citeauthor{Herlant.2016modeswitch} addresses these challenges by reducing the number of \acp{DoF} through mode switches. 
In their successful implementation, a joystick was used to control a \emph{Kinova Jaco} assistive robotic arm~\cite{Herlant.2016modeswitch}.



Alternatively, \citeauthor{Arevalo-Arboleda2021} introduced a hands-free multi-modal interaction by combining head movements, using a head-gaze based cursor to point, and speech commands to execute specific actions for tele-operating a robotic arm~\cite{Arevalo-Arboleda2021}. However, while speech commands provide enhanced accessibility, challenges like environmental noise or speech impairments encounter, impacting their effectiveness~\cite{Mayuri.2020EMIVR}.

The control of assistive robotic arms involves a wide array of possible input devices, each targeted to suit the preferences and capabilities of the respective user~\cite{Arevalo-Arboleda2021b}. Despite this diversity, there remains a gap in the evaluation of these input devices within the context of \ac{AI}-enhanced shared control applications for assistive robots.


In previous research, we introduced the \emph{AdaptiX} framework, an open-source \acs{XR} tool designed for \ac{DnD} operations~\cite{Pascher.2024adaptix}. \emph{AdaptiX} consists of a \ac{VR} simulation environment to prepare and test study settings as well as a \ac{ROS} interface to control a physical robotic arm. The framework also includes a general input adapter, facilitating the development and evaluation of different input technologies and devices. Leveraging these capabilities, \emph{AdaptiX} is used as the basis for this research project.

Through an algorithmic approach, the robotic arm's \acp{DoF} are configured to enable precise control with a low-\ac{DoF} input device. 
This adaptive \ac{DoF} mapping, denoted as \emph{\ac{ADMC}}, aims to present the user with a set of \ac{DoF} mappings, organized based on their effectiveness in executing the pick-and-place task employed in the experiment (\emph{optimal suggestion}, \emph{adjusted/orthogonal suggestion}, \emph{translation-only}, \emph{rotation-only}, and \emph{gripper}). 
The underlying concept of \enquote{usefulness} posits that optimizing the cardinal \acp{DoF} of the robot aligned with an input \ac{DoF} while advancing towards the next goal represents the most advantageous approach.

\section{Discrete and Continuous Control Methods}
Owning to \emph{AdaptiX}'s integration of \ac{ADMC}, users control the robotic arm forwards or backwards along a defined path based on the \ac{DoF} mapping. Consequently, only a single-\ac{DoF} input device is necessary for the movement. To choose from the different \ac{DoF} mapping suggestions of the system, an additional one-dimensional input is required to perform a \emph{mode switch} action, providing flexible and efficient control of the robotic arm.

Expanding upon the functionalities of \emph{AdaptiX}, this study focuses on discrete and continuous control methods serving as assistive input devices for the \ac{ADMC} shared control application. The framework's general input adapter provides a \emph{float} value ($-1.000$ -- $1.000$) for the \emph{Adaptive Axis} and a Boolean trigger for \emph{Switch Mode}.

\subsection{Motion Controller}
Prior studies~\cite{Pascher.2023inTimeAndSpace,Kronhardt.2022adaptOrPerish} used a \emph{Meta Quest} motion controller to interact with the \emph{AdaptiX} framework. 
To add to this, we integrated a \emph{Nintendo Joy-Con}~\cite{joycon}, which is well suited for one-handed operation. 
For the integration, we used \emph{UE4-JoyConDriver}~\cite{joyconControl} -- a plugin for \emph{Unreal Engine 4.27/5.2}.
The plugin creates a connection between \emph{Unreal} and \emph{Nintendo Joy-Con} and provides sensor data such as accelerometer, gyroscope and \acp{IMU}.

The left controller was selected for its balanced layout, accommodating both left- and right-handed users.
The thumbstick -- providing continuous data -- was tilted up or down to move the robotic arm forward or backward. The mode switch is performed by pressing the \emph{Up}-button of the controller right beneath the thumbstick. This design ensures single-handed control of the robot while preventing simultaneous movement and mode switching for enhanced usability.

\subsection{Head-based Control}
This control method eliminates the need for extra, specialized input devices as it utilizes orientation data from a device the user is already using -- the \ac{HMD}~\cite{varjo}. Furthermore, it offers an accessible approach by allowing users with impaired hand motor function to operate the robotic arm.

The \ac{HMD}'s internal sensor technology, specifically the \acp{IMU}, facilitates the measurement of head rotations along three axes (\emph{roll}, \emph{pitch}, and \emph{yaw}). This coordinate system is anchored to the object, positioned at the center of the user's head. 
Positive and negative rotations are possible around each axis, facilitating the mapping of six distinct actions to the corresponding axis rotations.

When the user tilts their head in a positive manner (\emph{pitch}; rotating the head upwards), the robotic arm is advanced along the \ac{DoF} mapped trajectory. Conversely, tilting the head in the opposite direction causes the arm to move backward along that path. 
Rolling the head to the right triggers the mode switch action, selecting the next \ac{ADMC} suggestion.

Along each head rotation axis, a 20\degree\ resting zone has been set to prevent unintentional controlling of the robot.
In this application, the user's head serves as a continuous data source for controlling the robot, akin to a joystick or the \emph{Joy-Con's} thumbstick.

\subsection{Assistive Buttons}
Integrating the \emph{Microsoft Xbox Adaptive Controller}~\cite{xboxcontroller}, emphasizing flexibility and accessibility, enables the use of assistive buttons (e.g., \emph{Logitech Adaptive Gaming Kit}~\cite{logitechkit}).
These can be quickly and flexibly arranged to ensure comfortable operation by the user.

Similar to a gamepad control for discrete input data, the elementary actions for moving forward and backwards are mapped onto the adaptive buttons. The buttons marked \emph{Arrow up} and \emph{Arrow down} are mapped for moving the robotic arm, while a button with an \emph{A}-marking was assigned to the mode switch.

\section{Study}
This preliminary study gathered initial user experiences with different modalities and operating modes for \ac{AI}-enhanced assistive robotic arms. Through a controlled \ac{MR} user study involving 14 participants (6 female, 8 male), we systematically compare the advantages and disadvantages of the selected input methods.
Four participants had prior experience with robotic arms. 


\subsection{Study Design}
We employed a within-participant experimental design, with the \emph{control method} as the independent variable, comprising three conditions: (1) \emph{Joy-Con}, (2) \emph{Head}, and (3) \emph{Button}. 
Each participant underwent eight trials per condition.
To mitigate the potential impacts of learning and fatigue, the condition order was fully counterbalanced.





\subsection{Apparatus}

\begin{figure}[htbp]
    \centering
    \includegraphics[width=\linewidth]{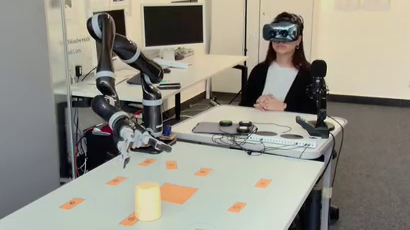}
    \caption{Overview of the study setup. The participant is wearing a \emph{Varjo XR-3} \ac{HMD} and controls the \emph{Kinova Jaco 2} via head movements. The goal is to grasp the light-colored rounded block and place it on the large orange square in the middle of the table. The small orange markings are potential starting points for the rounded block.}
    \label{fig:study-setup}
    \vspace{-4mm}
\end{figure}

Our study used the \emph{AdaptiX}~\cite{Pascher.2024adaptix} framework to integrate and assess the selected control methods. We operated the framework in its \ac{MR} mode, employing the \emph{Varjo XR-3}~\cite{varjo} \ac{HMD} and a \emph{Kinova Jaco 2}~\cite{kinova} assistive robotic arm, as shown in~\autoref{fig:study-setup}. We connected the \emph{Varjo XR-3} and all input devices to a \emph{Schenker Media Station} computer to facilitate this setup. Furthermore, we established connections between the \emph{Schenker Media Station}, the \ac{ROS} server, and the \emph{Kinova Jaco 2} through a wired \ac{LAN}.

\subsection{Procedure}

Before starting, participants received a detailed explanation of the project's objectives and the tasks involved. Each participant provided informed consent for their participation, including recording video, audio, and any other relevant data. A study administrator, overseeing the experiment on a laptop, provided instructions on using the hardware and the study environment. Once set up, participants followed command prompts within the \ac{MR} environment. For each of the three conditions, the following steps were performed: 

\begin{enumerate}
    \item Participants were given a written and standardized explanation of the control method used in the current condition.
    \item Participants conducted eight trials, grasping the object and placing it on the target surface.
    \item Interview and questionnaires.
\end{enumerate}

After completing all conditions, participants ranked the three control methods from \emph{most} to \emph{least preferred} and explained their decision. The study concluded with a de-briefing. 

\subsection{Experimental Task}
The experimental procedure builds on prior research that employed the \emph{AdaptiX} framework (refer to ~\cite{Pascher.2023inTimeAndSpace}). The present study expands the configuration to a real-world environment, replicating a typical pick-and-place scenario.

To commence each trial, the study administrator positioned an object on a table. The participant aimed to navigate the robot from its initial location to grasp the object and deposit it onto a designated target area on the same table. For each trial, the object's starting position varied among eight possible predetermined locations. These positions were randomized in their sequence. We employed uniform rounded block shapes as objects to ensure impartiality and trial comparability, eliminating bias and allowing for consistent trial comparisons.
Users could adjust the robot's \ac{DoF} mapping by toggling between modes to fulfill the task. Following a successful execution, the object was removed, and the robot returned to its initial position. The object was then placed in a new starting position for a subsequent trial to begin. 
Upon completing each condition, we assessed workload using the \ac{RTLX} questionnaire~\cite{Hart.2006} and measured the five dimensions of the \ac{QUEAD}~\cite{schmidtler2017questionnaire}. The task completion time was recorded from the moment the participant initiated the movement of the robotic arm until the block was successfully placed.



\section{Results}
This research focused on collecting subjective feedback from participants to improve the future development and integration of control input methods for shared control applications. The presented study encompasses a total of 336 (14 participants $\times$ 3 control methods $\times$ 8 trials) measured trials.


\subsection{Perceived workload}
\begin{figure}[htbp]
    \centering
    \includegraphics[width=1\linewidth]{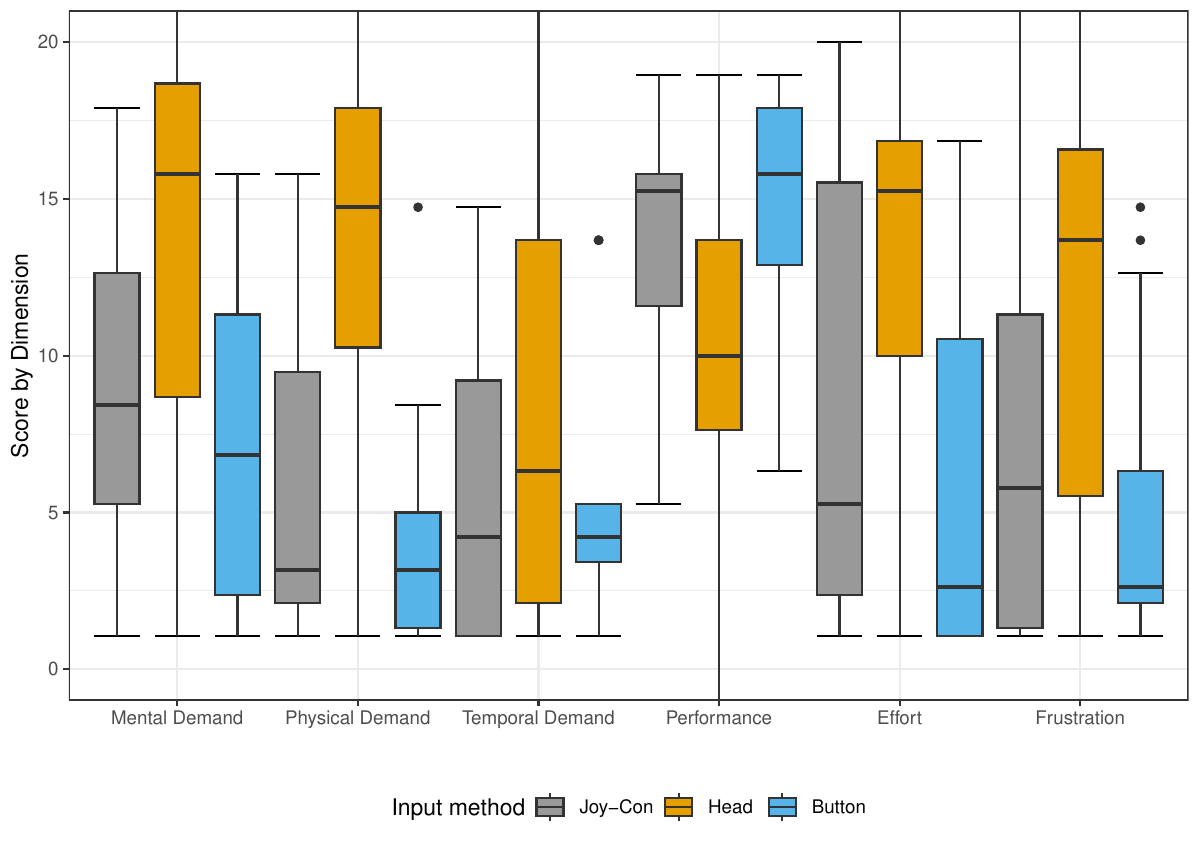}
    \caption{Comparison of the task load dimensions for the three different control methods: \textit{Joy-Con}, \textit{Head}, and \textit{Button}}
    \label{fig:NASA_TLX}
    \vspace{-4mm}
\end{figure}

\ac{RTLX}~\cite{Hart.2006} scores [scale from 1 to 100] for all participants resulted in mean task load values of \emph{Button}~$=$~35.90 (SD~$=$~12.98), \emph{Joy-Con}~$=$~41.17 (SD~$=$~18.66), and \emph{Head}~$=$~59.65 (SD~$=$~19.64). We applied a Friedman test which revealed a significant main effect for perceived task load ($\chi^2$(2)~$=$~18.00, p~$\leq$~0.001~***, N~$=$~14). The post-hoc pairwise comparisons (Bonferroni corrected) using Wilcoxon signed-rank tests revealed significant differences between \emph{Head} and \emph{Button} (Z~$=$~$-$3.27, p~$\leq$~0.001~***, r~$=$~0.67), \emph{Head} and \emph{Joy-Con} (Z~$=$~$-$3.02, p~$=$~0.002~**, r~$=$~0.62), but not between \emph{Button} and \emph{Joy-Con} (Z~$=$~$-$1.44, p~$=$~0.487, r~$=$~0.29).
The resulting task load scores per individual dimension of the \ac{RTLX} are presented in \autoref{fig:NASA_TLX}.

\subsection{Evaluation of Physical Assistive Devices}
The \ac{QUEAD} included five individual scales (7-point Likert). Friedman tests for individual dimensions revealed significant main effects for all dimensions. Post-hoc pairwise comparisons indicate significant differences between \emph{Head} and \emph{Button} for all five dimensions  as well as between \emph{Head} and \emph{Joy-Con} for \emph{PU}, \emph{PEU}, \emph{E}, and \emph{C}. For \emph{Joy-Con} and \emph{Button} only \emph{PU} and \emph{PEU} show significant differences (refer to Table~\ref{tab:quead} for detailed scores). 
\vspace{-1mm}
\begin{table}[htbp]
    \centering
    \captionsetup{justification=justified}
    \caption{Statistics for individual \ac{QUEAD} dimensions: Perceived Usefulness (PU), Perceived Ease of Use (PEU), Emotions (E), Attitude (A), and Comfort (C). \label{tab:quead}}
    \footnotesize
    \begin{tabular}{p{1.3cm}ccccc}
        \toprule
        \textbf{}	& \textbf{PU} & \textbf{PEU} & \textbf{E} & \textbf{A} & \textbf{C} \\
        \midrule
        \multicolumn{6}{c}{\textbf{Descriptive Statistics: mean value (standard deviation)}}                 \\
        \midrule
        \emph{Joy-Con}             & 4.84 (1.18)  & 5.11 (1.55)  & 5.19 (1.41)  & 4.46 (2.16)  & 5.71 (1.87) \\
        \emph{Head}                   & 2.93 (1.28)  & 3.37 (1.54)  & 2.86 (1.67)  & 2.82 (2.24)  & 3.57 (1.96)  \\
        \emph{Button}               & 5.63 (1.44)  & 6.06 (1.15) & 5.90 (1.22) & 5.61 (1.84) & 5.79 (1.46) \\
        \midrule
        \multicolumn{6}{c}{\textbf{Friedman Tests}}                         \\
        \midrule
        $\chi^2$(2)                   & 20.48       & 13.00             & 16.57         & 9.69      & 11.78      \\
        $p$                           & $\leq$0.001~*** & 0.002~**             & $\leq$0.001~***   & 0.008~**     & 0.003~**   \\
        $N$                           & 14          & 14                & 14            & 14        & 14        \\
        \midrule
        \multicolumn{6}{c}{\textbf{Pairwise Comparisons}}                                               \\
        \midrule
        \multicolumn{3}{l}{\emph{Joy-Con} vs. \emph{Head}}                            &           &           &           \\
        \midrule
        $|Z|$                         & 2.99           & 2.42              & 2.58      & 1.70       & 2.88      \\
        $p$                           & 0.004~**       & 0.040~*             & 0.023~*     & 0.282       & 0.006~**     \\
        $r$                           & 0.56           & 0.46              & 0.49      & 0.32       & 0.54      \\
        \midrule
        \multicolumn{3}{l}{\emph{Head} vs. \emph{Button}}                            &           &           &           \\
         \midrule
        $|Z|$                         & 3.27            & 3.17              & 3.11      & 3.02       & 2.97      \\
        $p$                           & $\leq$0.001~*** & 0.001~***             & 0.002~**     & 0.003~**       & 0.006~**   \\
        $r$                           & 0.62            & 0.60              & 0.59      & 0.57       & 0.56      \\
        \midrule
        \multicolumn{3}{l}{\emph{Joy-Con} vs. \emph{Button}}                           &           &           &           \\
         \midrule
        $|Z|$                         & 2.58      & 2.39              & 1.79      & 1.54       & 0.16      \\
        $p$                           & 0.023~*     & 0.044~*             & 0.227     & 0.382       & $\geq$0.999     \\
        $r$                           & 0.49      & 0.45              & 0.34      & 0.29       & 0.03      \\
        \bottomrule
    \end{tabular}
    \vspace{-1mm}
\end{table}

\subsection{Individual Ranking}
All participants -- except one -- ranked conditions from 1 = \emph{favorite} to 3 = \emph{least favorite}. 
Mean values in ascending order are \emph{Button}~$=$~1.46 (SD~$=$~0.52); \emph{Joy-Con}~$=$~1.69 (SD~$=$~0.63); and \emph{Head}~$=$~2.85 (SD~$=$~0.55). A Friedman test revealed a significant main effect ({$\chi^2$(2)}~$=$~14.31, p~$=\leq$0.001~***, N~$=$~13). The post-hoc pairwise comparisons indicate significant differences between \emph{Head} \& \emph{Button} (Z~$=$~3.02, p~$=$~0.005~**, r~$=$~0.59) and \emph{Head} \& \emph{Joy-Con} (Z~$=$~2.52, p~$=$~0.026~*, r~$=$~0.49), but not between \emph{Button} \& \emph{Joy-Con} (Z~$=$~$-$0.83, p~$\geq$0.999, r~$=$~0.16).


\subsection{Subjective Feedback}
Participants noted an increased mental workload during the \emph{Head}-based interaction. P01 highlighted that the movement execution for \enquote{forward felt opposite to the suggested arrow direction}. Additionally, P01 got quickly distracted by a conversation with the experimenter, and P02 required substantial assistance due to difficulties in perceiving the arrows and mapping them to the head-movement direction.
Participants P01 -- P04 suggested introducing an additional mode switch to display the previous suggestion rather than presenting the next one. Participants P04, P11, and P12 preferred a non-continuous control by moving the head (i.e., only stop and go) to \enquote{prevent unintentional robot control when returning their head to the zero position} (P11). 

Similar to the \emph{Head}-based interactions, participants P01 -- P04 mentioned a discrepancy between the suggested arrows by the system and the control input. In certain situations, the system suggests movements in the user's direction. To move the robot along this trajectory (\emph{forwards}), the thumbstick of the \emph{Joy-Con} or \emph{Arrow Up} assistive button had to be pressed, which felt \enquote{discrepant}. 
Participant P04 suggested using the thumbstick of the \emph{Joy-Con} instead of the selected button for mode switching, for example, by tilting it sidewards.

Additionally, it was observed that specific initial placements of the object were perceived as disadvantageous compared to others, as the robot is fixed in place and has to perform -- for the novice users -- un-legible movements to reach the target.
\section{Discussion}
All participants were able to control the robotic arm with each input device to fulfill the project task. Yet, the study's findings indicate that the effectiveness of the \emph{Head}-based interaction method for controlling the robotic arm is relatively low compared to both hand-operated input methods. A notable insight derived from these results is the potential issue of the \emph{Varjo XR-3} \ac{HMD} being too bulky and heavy for sustained and precise \emph{Head}-based control. To address these concerns, a more lightweight and comfortable solution, such as utilizing external \acp{IMU} for \emph{Head}-based interaction~\cite{Wohle.2018marg,Jackowski.2018imu}, could be considered. 

Nevertheless, the \ac{HMD} remains essential for visualizing directional cues, even with the integration of \acp{IMU}. Looking forward, advancements in technology are expected to yield significantly more compact and lighter devices, thereby enhancing user comfort and immersion.



Further, participants pointed out a discrepancy between the robot's movement direction and the mapping of user inputs. This could lead to an unclear mental model, particularly since the robot is controlled in a \emph{first-person view}. To counteract this issue, a more extensive familiarization phase might be beneficial.
\section{Conclusion}

The input methods \emph{Joy-Con} and \emph{Button} represent promising approaches for controlling a robotic arm in a shared control application. Notably, both hand-operated input methods -- irrespective of whether they provide discrete or continuous input data -- (1) reduced perceived user workload and (2) improve \emph{Perceived Usefulness}, \emph{Perceived Ease of Use}, \emph{Emotions}, and \emph{Comfort}. 
These findings hold valuable implications for \ac{HRI} researchers involved in the design of input technologies for assistive robotic arms. Future research efforts should prioritize the nuanced analysis of both quantitative and qualitative feedback obtained from focus groups. This comprehensive approach aims to refine and develop optimal methods for robot motion control, with the overarching goal of improving usability, safety, and end-user acceptance of these technologies.

Still, given the diverse likes and dislikes of the participants, future development of adaptive input control methods should -- in line with \citeauthor{Burkolter.2014customization} -- include individualization options to increase comfort and end-user acceptance~\cite{Burkolter.2014customization}.

\begin{acks}
This research is supported by the \textit{German Federal Ministry of Education and Research} (BMBF, FKZ: \href{https://foerderportal.bund.de/foekat/jsp/SucheAction.do?actionMode=view&fkz=16SV8565}{16SV8565}). Our study is approved by the Ethics Committee of the \textit{Faculty of Business Administration and Economics of the University of Duisburg-Essen}.
\end{acks}

\bibliographystyle{ACM-Reference-Format}
\balance
\bibliography{MainPaper}

\end{document}